\documentclass[pdflatex,sn-mathphys-num]{sn-jnl}% Math and Physical Sciences Numbered Reference Style
\usepackage{graphicx}%
\usepackage{multirow}%
\usepackage{amsmath,amssymb,amsfonts}%
\usepackage{amsthm}%
\usepackage{mathrsfs}%
\usepackage[title]{appendix}%
\usepackage{xcolor}%
\usepackage{textcomp}%
\usepackage{manyfoot}%
\usepackage{booktabs}%
\usepackage{algorithm}%
\usepackage{algorithmicx}%
\usepackage{algpseudocode}%
\usepackage{listings}%
\usepackage{siunitx}
\usepackage{tabularray}
\UseTblrLibrary{booktabs}
\usepackage{tikz}
\usetikzlibrary{arrows.meta, positioning, fit, backgrounds, calc, shapes.geometric, shadows.blur}
\usepackage{fontawesome7}
\usepackage[capitalise,noabbrev,nameinlink]{cleveref}
\usepackage[colorinlistoftodos,prependcaption,textsize=tiny]{todonotes}

\usetikzlibrary{
  arrows.meta,
  positioning,
  shapes.misc,
  decorations.pathreplacing,
  patterns
}

\definecolor{signalblue}{HTML}{5B9BD5}
\definecolor{safecondition}{HTML}{70AD47}
\definecolor{riskycondition}{HTML}{FFC000}
\definecolor{successgreen}{HTML}{00B050}
\definecolor{failurered}{HTML}{C00000}
\definecolor{infogray}{HTML}{808080}
\definecolor{zonecolor}{gray}{0.9}
\definecolor{layercolor}{gray}{0.6}
\definecolor{gridcolor}{gray}{0.9}

\tikzset{
  timeline/.style={-Latex, thick},
  layer_label/.style={font=\bfseries, color=layercolor},
  event/.style={circle, fill=signalblue, inner sep=2.5pt, text=white},
  zone/.style={rectangle, rounded corners=2pt, fill=riskycondition, fill opacity=0.15, draw=riskycondition, inner ysep=1mm},
  anno/.style={font=\small, text=infogray, align=center},
  anno_success/.style={anno, text=successgreen},
  anno_failure/.style={anno, text=failurered},
  icon_success/.style={font=\large, text=successgreen},
  icon_failure/.style={font=\large, text=failurered},
  status_box/.style={rounded rectangle, draw, thin, font=\small, align=center},
  ok_box/.style={status_box, fill=successgreen!15, draw=successgreen},
  error_box/.style={status_box, fill=failurered!15, draw=failurered},
}

\usepackage{listings}
\usepackage{xcolor}
\lstdefinelanguage{EPL}{
  morekeywords={SELECT,INSERT,INTO,FROM,WHERE,GROUP,BY,HAVING,AS,cast,payload,as,CREATE,SCHEMA,PATTERN,EVERY,UNTIL,NOT,WITHIN,OUTPUT,ALL,EVENTS,SECONDS,SECOND,SEC,WIN,TIME,ISTREAM,RETAIN,FIRST,LAST,PREV,COALESCE,CASE,WHEN,THEN,END,OR,AND,BETWEEN,IS,NULL,LIKE,REGEX,LIMIT,ANNOTATION,WINDOW,UPDATE,DELETE,MERGE,PATTERN},
  sensitive=true,
  literate={->}{{\textcolor{blue!60!black}{$\rightarrow$}}}1,
  morecomment=[l]{--},
  morestring=[b]'
}
\lstdefinestyle{epl}{
  float=t,
  basicstyle=\ttfamily\scriptsize,
  keywordstyle=\bfseries\color{blue!60!black},
  commentstyle=\itshape\color{gray!70},
  stringstyle=\color{green!40!black},
  showstringspaces=false,
  frame=single, framerule=0.3pt, rulecolor=\color{black!15},
  breaklines=true, tabsize=2,
}

\usepackage{enumitem}
\setlist[itemize,enumerate]{noitemsep,topsep=1ex}
\theoremstyle{thmstyleone}%
\theoremstyle{thmstyletwo}%

\theoremstyle{thmstylethree}%

\begin{document}

\title[Hybrid Declarative Process Execution via CEP]{Executing Discrete/Continuous Declarative Process Specifications via Complex Event Processing}

%%=============================================================%%
%% GivenName  -> \fnm{Joergen W.}
%% Particle  -> \spfx{van der} -> surname prefix
%% FamilyName  -> \sur{Ploeg}
%% Suffix  -> \sfx{IV}
%% \author*[1,2]{\fnm{Joergen W.} \spfx{van der} \sur{Ploeg}
%%  \sfx{IV}}\email{iauthor@gmail.com}
%%=============================================================%%

\author*[1]{\fnm{Leo} \sur{Poss}}\email{poss.leo@ur.de}
\author[1]{\fnm{Stefan} \sur{Schönig}}\email{stefan.schoenig@ur.de}
\author[2]{\fnm{Fabrizio Maria} \sur{Maggi}}\email{maggi@inf.unibz.it}

\affil*[1]{\orgname{University of Regensburg}, \orgaddress{\city{Regensburg}, \country{Germany}}}
\affil[2]{\orgname{Free University of Bozen-Bolzano}, \orgaddress{\city{Bolzano} \country{Italy}}}

\abstract{Traditional Business Process Management (BPM) focuses on discrete events and fails to incorporate critical continuous sensor data in cyber-physical environments. Hybrid declarative specifications, utilizing Signal Temporal Logic (STL), address this limitation by allowing constraints over both discrete events and real-valued signals. However, existing work has been limited to monitoring and post-hoc conformance checking. This paper introduces a novel execution architecture based on Complex Event Processing (CEP) that enables the real-time execution and enforcement of hybrid declarative models. Our three-layer approach integrates STL-inspired predicates into the execution flow, allowing the system to actively trigger activities and enforce process boundaries based on continuous sensor behavior. This approach bridges the gap between hybrid specification and operational control.}


\maketitle

\section{Introduction}

Business Process Management (BPM) has traditionally focused on modeling, monitoring, and optimizing processes based on \emph{discrete event data}, i.e., sequences of activities representing the execution of business processes~\cite{dumas2018fundamentals}. Declarative approaches such as \emph{Declare}~\cite{pesic2007declare,burattin2016conformance} have emerged as a powerful alternative to imperative modeling languages, shifting the focus from explicitly describing control flow to specifying \emph{constraints} that must hold during process execution. This paradigm is particularly valuable in dynamic and flexible environments, where not all execution paths can be anticipated in advance~\cite{Maggi19}. However, modern digital systems increasingly operate in \emph{cyber-physical} or \emph{data-rich} environments, where discrete events alone no longer capture the full state of a process. In domains such as \emph{Industry 4.0}~\cite{bazan2022industry} or \emph{Internet of Things (IoT)-enabled healthcare}~\cite{janiesch2020iotbpm}, processes unfold alongside continuous sensor data (e.g., temperature, pressure, machine vibrations, or patient vital signs) that directly influence process decisions. Traditional process models, including classical Declare specifications, cannot directly reason about such continuous signals, as they were designed for discrete traces only. Consider a chemical reactor in a smart factory. A process engineer specifies a safety boundary: if the reactor temperature stays above 80\,°C for more than 10 consecutive seconds, the activity \texttt{StartCooling} must be executed within 5 seconds to prevent thermal runaway. Standard declarative workflows cannot express or enforce this rule as they lack continuous-time semantics and cannot reason about durations in real-valued sensor streams. Classical data-aware process extensions fare no better: they operate on static data snapshots at task boundaries and cannot detect signal fluctuations between tasks.

Recent work has begun to address this limitation by employing \emph{Signal Temporal Logic (STL)}~\cite{maler2004monitoring}, a formalism from the cyber-physical systems domain that allows the specification of temporal properties over real-valued signals. STL enables constraints such as \emph{temperature must never exceed 40\,°C} or \emph{if pressure rises above 20\,bar, a safety valve must open within 10\,seconds.} Building on STL, researchers have proposed \emph{hybrid declarative models}\footnote{The term \emph{hybrid} in this context has been introduced in \cite{cimatti2009requirements}.} that unify discrete events and continuous data streams by encoding discrete events as boolean signals and combining them with continuous predicates~\cite{corea2025hybrid}. This enables \emph{conformance checking} of hybrid traces (including both discrete and continuous information) to verify whether observed executions satisfy these hybrid specifications. While this represents a significant step forward, these approaches remain limited to \emph{monitoring} and \emph{post-hoc conformance checking}~\cite{donze2010robust,burattin2016conformance}. Hybrid declarative models have proven effective for verifying logs and sensor data \emph{after execution} or for runtime monitoring; however, no approach currently supports their \emph{execution}, i.e., using these specifications to actively drive and control process instances in real time.

In this paper, we address this gap. Building on our previous work on executing declarative specifications via Complex Event Processing (CEP)~\cite{Poss2025_Synergistic}, we extend this approach to support \emph{hybrid declarative models} that integrate continuous sensor data with discrete event streams. Our approach introduces a signal-processing layer that transforms continuous data into real-time predicates and integrates them into a CEP-based execution engine. This enables the specification and enforcement of STL-like constraints during runtime, for example, triggering activities when continuous conditions hold or proactively enforcing process boundaries based on sensor behavior.
In summary, the main contributions of this work are:
\begin{itemize}
  \item We extend the execution paradigm of declarative process models from purely discrete event data to \emph{hybrid event–signal} scenarios.
  \item We present a \emph{CEP-based execution architecture} capable of evaluating STL-inspired predicates and constraints in real time.
  \item We show how hybrid declarative constraints can be \emph{enforced during execution}, enabling reactive and proactive behavior in data-rich environments.
\end{itemize}

This work bridges the gap between the specification of hybrid declarative constraints and their execution, enabling the development of next-generation process-aware systems that are both \emph{context-sensitive} and \emph{sensor-integrated}.

The remainder of this paper is structured as follows. \Cref{sec:fundamentals} covers declarative modeling, STL, and the principles of CEP as a stream reasoning paradigm. \Cref{sec:concept} introduces our core contribution: the three-layer hybrid execution architecture and an implementation-independent mapping of hybrid constraints to CEP logic. \Cref{sec:implementation} describes the working prototype built on the Esper engine. \cref{sec:evaluation} assesses expressiveness and scalability. \Cref{sec:relatedwork} discusses related work, and \cref{sec:outlook} concludes with future research directions.

\section{Preliminaries and Fundamentals}\label{sec:fundamentals}

This section provides the theoretical foundation for our work. We first discuss the fundamentals of temporal logic and declarative process modeling based on LTL. We then present STL and explain its role in enabling hybrid declarative specifications that integrate discrete events with continuous data.

\subsection{Linear Temporal Logic and Declarative Process Modeling}

Linear Temporal Logic (LTL)~\cite{pnueli1977temporal} is a classical formalism for specifying temporal properties of discrete-event systems. LTL formulas are evaluated over \emph{linear traces}, i.e., sequences of states ordered in time, and are used to specify constraints concerning the ordering and occurrence of events. Given a set of atomic propositions $AP$, the syntax of LTL is defined as:
\[
  \varphi ::= \top \mid p \mid \neg \varphi \mid \varphi_1 \land \varphi_2 \mid \mathsf{X}\varphi \mid \mathsf{F}\varphi \mid \mathsf{G}\varphi \mid \varphi_1 \, \mathsf{U} \, \varphi_2
\]
where $p \in AP$, $\mathsf{X}$ denotes ``next'', $\mathsf{F}$ ``eventually'', $\mathsf{G}$ ``globally'', and $\mathsf{U}$ ``until''. For example, the formula $\mathsf{G}(a \rightarrow \mathsf{F} b)$ expresses that ``whenever $a$ occurs, $b$ must eventually occur''. Such expressions have been widely used for verifying temporal properties and form the logical basis for many declarative approaches in business process management~\cite{Maggi19}.

Declarative process modeling builds on these principles and shifts the perspective from explicitly modeling control flow to defining a set of \emph{constraints} that must hold during process execution. One of the most widely adopted declarative languages is \emph{Declare}~\cite{pesic2007declare}, whose templates are typically defined as LTL (more specifically LTL over finite traces, LTL$_f$) formulas. Each Declare constraint captures a behavioral rule that restricts the space of acceptable traces. Examples of commonly used templates include:

\begin{itemize}
  \item \textsc{Response}$(a,b)$: $\mathsf{G}(a \rightarrow \mathsf{F} b)$, stating that whenever $a$ occurs, $b$ must eventually occur.
  \item \textsc{NotExistence}$(a)$: $\mathsf{G}(\neg a)$, specifying that $a$ must never occur.
  \item \textsc{Precedence}$(a,b)$: $(\neg b \, \mathsf{U} \, a) \vee \mathsf{G}(\neg b)$, meaning that $b$ may occur only after $a$ has occurred.
\end{itemize}

This modeling style offers a high degree of flexibility and is particularly suited to dynamic, knowledge-intensive processes~\cite{Maggi19}. It has been extensively used for runtime monitoring, conformance checking, and process discovery~\cite{Maggi19}. Beyond pure control-flow constraints, Declare has been extended to \emph{multi-perspective} (MP) specifications through Metric First-Order Temporal Logic (MFOTL)~\cite{burattin2016conformance}. MFOTL enriches LTL with first-order quantification over data variables (e.g., event payloads) and metric time intervals, enabling constraints such as \texttt{Response(A, B)[A.orderId = B.orderId]} that correlate activation and target events based on their data attributes. Our architecture builds on this MP-Declare foundation, further extending it with continuous signal predicates drawn from STL.

Nevertheless, LTL and Declare, even in their multi-perspective form, share a fundamental limitation: they operate over discrete, symbolic event logs and cannot naturally express conditions involving real-valued data or continuous-time behavior. In many cyber-physical or IoT-enabled scenarios, such as smart manufacturing or healthcare, process decisions depend on continuous sensor streams (e.g., temperature, pressure, oxygen saturation). Capturing such dependencies requires a more expressive temporal logic capable of reasoning over real-valued signals.

\subsection{Signal Temporal Logic and Hybrid Declarative Specifications}

STL~\cite{maler2004monitoring} extends temporal reasoning beyond discrete events to encompass real-valued, time-continuous signals. STL formulas are evaluated over signals $x: \mathbb{R}_{\geq 0} \rightarrow \mathbb{R}$, enabling the specification of temporal constraints involving thresholds and timing intervals. The syntax of STL is defined as:
\[
  \varphi ::= \top \mid x(t)\, op \, c \mid \neg \varphi \mid \varphi_1 \land \varphi_2 \mid
  \mathsf{F}_{[a,b]}\varphi \mid \mathsf{G}_{[a,b]}\varphi \mid
  \varphi_1 \, \mathsf{U}_{[a,b]} \, \varphi_2,
\]
where $op \in \{<,\leq,=,\geq,>\}$ and $[a,b]$ denotes a time interval. STL also supports \emph{robust semantics}~\cite{donze2010robust}, which quantify how strongly a signal satisfies or violates a property, thereby enabling tolerance to noise and uncertainty.

To reason simultaneously about discrete events and continuous data, we adopt the hybrid model construction introduced by Corea et al.~\cite{corea2025hybrid}. We define the set of state variables as a pair $V=(D,C)$, where $D$ is a set of discrete variables derived from event logs and $C$ is a set of continuous variables derived from data streams (e.g., IoT sensor measurements). From a standard event log, we extract a \emph{timed trace}, $tr = \langle (d_1,t_1), \dots, (d_n,t_n)\rangle$,
where $(d_i,t_i)$ denotes the occurrence of event $d_i \in D$ at time~$t_i$.

Since discrete events lack a continuous representation, we can transform each $d\in D$ into an \emph{augmented boolean signal} $S_d : T \rightarrow \{0,1\}$, set to~1 whenever $d$ is the most recent observed event, and to~0 otherwise. For notational convenience, we write $d[t]=S_d(t)$ to denote its value at time $t$. Continuous variables $c\in C$ are taken directly from their associated data streams, each producing values over the same continuous time domain $T=[0,m]\subset\mathbb{R}$. These components are then combined into a \emph{hybrid continuous trace} $\omega : T \rightarrow \mathbb{R}^{|D|+|C|}$, assigning to each variable (discrete or continuous) its value at each time instant.

Based on this representation, the STL fragment we use in this work is defined by the grammar:
\[
  \varphi ::= \mathbf{true}
  \mid c[t]\, op \, k
  \mid \mathsf{dis}(d[t])
  \mid (\neg \varphi)
  \mid (\varphi_1 \wedge \varphi_2)
  \mid (\varphi_1 \, \mathsf{U}_{I} \, \varphi_2),
\]
where $c\in C$, $d\in D$, $op\in\{<,\leq,=,\geq,>\}$, $k\in\mathbb{Q}$, and $I$ is a time interval in $T$.
The term $c[t]\, op \, k$ expresses constraints over continuous variables (e.g., ``the temperature must be larger than 100''), while $\mathsf{dis}(d[t])$ refers to the boolean signal associated with discrete event~$d$ at time $t$. 
Standard Declare templates lift to hybrid constraints through STL semantics, for example \textit{(i)} \textsc{Hybrid Response} that says whenever condition $\Theta_1$ holds, $\Theta_2$ must follow within 5 seconds,
      $
      \mathsf{G}_{[0,m]}(\Theta_1 \rightarrow \mathsf{F}_{[0,5]}\Theta_2).
      $ \textit{Example}: ``Whenever the temperature exceeds 80\,°C, the cooling activity must start within 5 seconds'', where
      $
      \Theta_1 := temp(t) > 80$, $\Theta_2 := \mathsf{dis}({StartCooling}[t])$, and
\textit{(ii)} \textsc{Hybrid NotExistence,} where a condition $\Theta$ must never hold continuously for more than duration $d$: $\neg \mathsf{F}_{[0,m]}(\mathsf{G}_{[0,d]}\Theta)$.
      \textit{Example}: ``Temperature must never exceed 90\,°C for more than 10 seconds, where $\Theta := temp(t) > 90$ and $d=10$\,s. Consider a trace of $m = 25$\,s over one continuous variable $C = \{\textit{temp}\}$ and two discrete activities $D = \{\texttt{StartCooling}, \texttt{Restart}\}$\footnote{The \texttt{Restart} task is not executed in the trace but will be relevant further down.}:
  \begin{itemize}
      \item $[0,10)$\,s: $\textit{temp}(t) = 75^\circ\text{C}$, no activity.
      \item $t = 10$\,s: temperature rises to $85^\circ\text{C}$ and holds.
      \item $t = 12$\,s: temperature spikes further to $93^\circ\text{C}$ and remains there.
      \item $t = 14$\,s: $\texttt{StartCooling}$ activity finishes.
      \item $t = 18$\,s: temperature drops below $80^\circ\text{C}$ (to $78^\circ\text{C}$) and remains steady.
  \end{itemize}

  \noindent The trace defines $\omega(t) = [\textit{temp}(t), S_{\texttt{StartCooling}}(t)]$ where $S_{\texttt{StartCooling}}(t) = 1$ for $t \ge 14.0$ and $0$ otherwise. The temperature exceeds $90^\circ$C continuously from $t = 12$\,s to $t = 18$\,s. Evaluated against the two constraints:
  \begin{enumerate}
      \item \textbf{Hybrid Response ($\Phi_2$), fulfilled:} $\textit{temp}(t) > 80$ activates at $t = 10.0$\,s, opening window $[10.0, 15.0]$. $S_{\texttt{StartCooling}}(t)$ becomes $1$ at $t = 14.0$\,s (inside the window) so the constraint is \textbf{fulfilled}.
      \item \textbf{Hybrid NotExistence ($\Phi_1$), fulfilled:} the invariant $\neg \mathsf{F}_{[0,25]}(\mathsf{G}_{[0,10]}(\textit{temp}(t) > 90))$ watches the stream. The temperature exceeds $90^\circ$C continuously from $t = 12$\,s, but drops at $t = 18$\,s, meaning it remained above $90^\circ$C for a continuous duration of only 6 seconds. Since this does not reach the 10-second duration threshold, the engine evaluates this constraint as \textbf{not violated}.
  \end{enumerate}

  \noindent Hybrid declarative specifications combine discrete events and continuous signals in a single constraint model---a capability beyond traditional Declare. Existing work~\cite{corea2025hybrid} applies these specifications to offline monitoring and conformance checking, but stops at post hoc verification. Our architecture extends them to runtime execution.

  \subsection{Complex Event Processing}
  Complex Event Processing (CEP) is a design paradigm engineered for processing high-velocity, continuous streams of real-time event data to identify actionable patterns, causal relationships, and temporal trends across multiple data sources~\cite{Luckham2008_Power,wu2006high,janiesch2020iotbpm}. In contrast to traditional Database Management Systems (DBMS) that execute momentary queries over static, persistent data tables, a CEP engine maintains static, continuous queries over dynamic data streams.

  Formally, an event stream can be defined as $S = \langle e_1, e_2, e_3, \dots \rangle$, an infinite sequence of event instances. Each individual event $e_i = (a, \vec{v}, t)$ is a tuple containing an event type identifier $a$, a vector of multi-perspective payload data attributes $\vec{v}$, and a discrete timestamp $t \in \mathbb{R}_{\geq 0}$ denoting its time of occurrence. A \emph{complex event} is a higher-level, semantically enriched event derived by the CEP engine when a set of primitive stream elements satisfies specific structural, relational, or temporal conditions. This pattern-matching mechanism typically uses sliding time windows ($W$) that bound the evaluation domain to a finite historical interval $\Delta t$.

  While the conceptual foundations of stream reasoning remain implementation-independent, operationalizing these paradigms requires a concrete execution engine. For this work, the \emph{Esper} CEP framework\footnote{\url{https://www.espertech.com/esper/}} was selected as the underlying infrastructure. Esper provides several distinct architectural advantages suited to declarative process enforcement: First, Esper's Event Processing Language (EPL) \cite{Langhi2025} extends standard SQL semantics with native regular expressions for event streams (via the \texttt{PATTERN} operator) and sequential causality matching using the linear-time operator (\texttt{->}). This maps directly to the forward- and backward-looking temporal structures of declarative templates. Second, it provides out-of-the-box support for programmatic time windows (\texttt{timer:within}, \texttt{timer:interval}), allowing the system to track real-time clock expirations and automatically emit complex negative events (such as timeouts). And finally, as Esper is built as an embedded Java library, it avoids the heavy network serialization overhead found in distributed stream frameworks (such as Apache Flink\footnote{\url{https://flink.apache.org/}} or Kafka Streams\footnote{\url{https://kafka.apache.org/43/streams/}}).\footnote{For additional context on the theoretical distinctions between Stream Processing and Complex Event Processing, particularly the difference between data-flow-oriented models (e.g., Flink, Storm) and pattern-detection-oriented models (e.g., Esper, Drools Fusion), see \cite{Cugola2012}, which offers a core taxonomy of the field, and \cite{Dayarathna2018}, which provides a more recent survey covering distributed CEP architectures and their scalability trade-offs. On the specific question of expressiveness in event pattern languages, \cite{Artikis2017} contrasts declarative (logic-based) and imperative (automaton-based) CEP approaches, a distinction directly relevant to Esper's hybrid EPL design discussed above.}

\section{Hybrid Declarative Process Execution}\label{sec:concept}

Corea et al.~\cite{corea2025hybrid} introduced a framework for specifying hybrid declarative constraints in STL, but limited to offline monitoring and post-hoc conformance checking over historical logs. We extend this to active runtime execution and prescriptive operational control based on standard CEP. In our architecture, hybrid Declare specifications drive a real-time process engine that shapes execution in two ways: \textit{(i)} when an STL temporal window expires, the engine provides the ability to implemented automated task execution (e.g., triggering a cooling valve); and \textit{(ii)} the engine evaluates the current trace state to enable, disable, or block human tasks at the interface providing a task list similar to other process execution engines.

  For complex interactions, data dependencies, or potentially conflicting multi-perspective constraints, the engine evaluates each rule independently based on its compiled CEP logic. At this stage of development, the architecture does not enforce an internal prioritization or resolution scheme for opposing actions. If a process engineer deploys a set of contradictory constraints (e.g., a rule forcing a reactive activity while an invariant simultaneously forbids it), the engine will evaluate both streams independently. This naturally results in an unfinishable or deadlocked process instance. We treat these operational conflicts strictly as design-time modeling failures (i.e., a flawed process specification) that have to be handled \textit{before} process execution. To realize this execution model, the architecture is organized as a three-layer runtime, where each layer performs a distinct transformation, progressively abstracting raw signals into an actionable, stateful process view:

  \begin{enumerate}
    \item Detection of activation and target events from \textit{atomic events} for each constraint, which then trigger another event at a higher abstraction level.
    \item Stateless management of the current constraint status using \textit{constraint-level events} for activation, target, violation, and additional basic events like process start and end. This layer is also required to detect interruptions.
    \item The current status of each constraint, i.e., process status, which, combined with the current trace, implies available tasks at the \textit{process level}.
\end{enumerate}

\subsection{Layered Architecture}
The execution architecture is organized as a three-layer runtime, where each layer performs a distinct transformation, progressively abstracting raw real-time streams into an actionable, stateful process view. The complete system pipeline is summarized in \cref{fig:architecture}.

\begin{figure}[t]
    \centering
    \begin{tikzpicture}[scale=.8, transform shape,
        font=\sffamily\small,
        block/.style={rectangle, draw=black!80, fill=white, text width=2.3cm, minimum height=0.9cm, align=center, rounded corners=3pt, thick, font=\sffamily\scriptsize},
        blockwide/.style={rectangle, draw=black!80, fill=white, text width=3.0cm, minimum height=0.9cm, align=center, rounded corners=3pt, thick, font=\sffamily\scriptsize},
        source/.style={rectangle, draw=black!70, fill=black!5, text width=1.6cm, minimum height=0.7cm, align=center, rounded corners=2pt, thick, font=\sffamily\scriptsize\bfseries},
        layer/.style={rectangle, draw=black!25, dashed, rounded corners=4pt, fill=black!2, thick},
        feedback/.style={-{Stealth[scale=0.95]}, dashed, thick, draw=failurered},
        line/.style={-{Stealth[scale=0.95]}, thick, draw=black!70},
      ]
      \filldraw[layer] (-1.8, 3.6) rectangle (5.2, 6.7);
      \filldraw[layer] (-1.8, 1.6) rectangle (5.2, 3.4);
      \filldraw[layer] (-1.8, -0.5) rectangle (5.2, 1.2);

      \node[anchor=east, font=\sffamily\bfseries\scriptsize, text=black!60, align=right] at (-2.0, 5.15) {L1: ATOMIC LAYER\\(Raw Ingestion)};
      \node[anchor=east, font=\sffamily\bfseries\scriptsize, text=black!60, align=right] at (-2.0, 2.5) {L2: CONSTRAINT LAYER\\(Complex Event Processing)};
      \node[anchor=east, font=\sffamily\bfseries\scriptsize, text=black!60, align=right] at (-2.0, 0.35) {L3: PROCESS LAYER\\(Aggregation \& Visualization)};

      \node[source] (events) at (0, 5.65) {Discrete Events};
      \node[source] (signals) at (3.2, 5.65) {Continuous Signals};
      \node[block] (l1_match) at (0, 4.45) {Event Matching\\ \& Signature Extraction};
      \node[block] (l1_signal) at (3.2, 4.45) {STL Predicate\\Evaluation};

      \node[blockwide] (l2_epl) at (0, 2.5) {Esper EPL Engine\\ $\text{LTL}_f/\text{STL} \rightarrow \text{EPL}(\varphi_a, \varphi_c, \varphi_t, \tau)$};
      \node[block] (l2_lifecycle) at (3.2, 2.5) {Constraint Lifecycle\\ \textsc{Init} $\rightarrow$ \textsc{Temp-Viol} $\rightarrow$ \textsc{Fulf}};

      \node[blockwide] (l3_analyzer) at (0, 0.35) {\textbf{AnalyzerService}\\Task Enabling \& Finishability};
      \node[block] (l3_dashboard) at (3.2, 0.35) {Task List and Dashboard};

      \node[block, draw=failurered, fill=white, text width=3cm] (enforcement) at (7.1, 2.5) {\textbf{Enforcement}-\textbf{Service}\\Dual-Rule Engine\\($\Sigma$, $\pi$, $\delta$, $\alpha$, $\rho$, $\gamma$)};

      \draw[line] (events) -- (l1_match);
      \draw[line] (signals) -- (l1_signal);
      \draw[line] (l1_match) -- (l2_epl);
      \draw[line] (l1_signal) -- ($(l2_epl.north east)!0.3!(l2_epl.north)$);
      \draw[line] (l2_epl) -- (l2_lifecycle);
      \draw[line] (l2_lifecycle) -- (l3_dashboard);
      \draw[line] (l2_epl) -- (l3_analyzer);
      \draw[line] (l3_analyzer) -- (l3_dashboard);

      \draw[line] (l1_signal.east) -| (enforcement.north);
      \draw[line] (l2_lifecycle.east) -- (enforcement.west);

      \draw[feedback] (enforcement.south) -- ++(0,-0.4) -- ++(2.1, 0) |- (0, 7.1) -| (-2.2, 5.65) -- (events.west);
    \end{tikzpicture}
    \caption{DeclareCEP system architecture: raw task events and continuous signals are ingested at L1, processed through EPL-encoded MP-Declare templates at L2 with duration enforcement, and exposed to state aggregation and UI monitoring at L3.}
    \label{fig:architecture}
    \vspace{-1\baselineskip}
\end{figure}

\paragraph{Atomic Event Layer (L1)}
The bottom layer, referred to as the \emph{Atomic Event Detection Layer (L1)}, is responsible for normalizing input streams and transforming them into semantically meaningful data tokens. The underlying stream infrastructure handles two primary input types: (i) discrete \emph{atomic events}, representing the completion of business activities, and (ii) continuous \emph{signal samples}, representing time-stamped telemetry measurements. Each atomic event is described as a tuple $e = \langle caseId,\, activity,\, ts \rangle$, representing activity occurrences such as \texttt{PickItem} or \texttt{Ship}, with each active task tracked as a boolean signal $d[t]$. Continuous input is represented as $s = \langle caseId,\, sensorId,\, value,\, timestamp \rangle$, and evaluated against continuous predicates of the form \( c[t] \; op \; k \) in real time. 

In contrast to post-hoc monitoring approaches which convert historical logs into augmented boolean signals after execution, this runtime engine performs hybridization continuously. Relational filtering predicates evaluate incoming metrics against static signal boundaries (e.g., comparing a telemetry sample value against a defined threshold) to broadcast binary state-change tokens. These structural tokens are explicitly categorized as \texttt{ACTIVATION} or \texttt{TARGET} signals and emitted into a consolidated intermediate stream along with their parent constraint identifier. This ensures that the engine is not forced to store massive raw signal traces in memory.

\paragraph{Constraint Status Layer (L2)}
The second layer of the runtime, the \emph{Constraint-Level Pattern Matching Layer (L2)}, is a stateless stream processor that detects temporal patterns defining a constraint's logic. It consumes the stream of standardized activation and target tokens from \textbf{L1} and maps sequential lifecycle shifts through pattern queries. For each constraint template, this layer deploys stream queries that look for specified sequential and metric dependencies between \texttt{ACTIVATION} and \texttt{TARGET} elements over sliding operational windows ($\Delta t$). For example, a \textsc{Response} constraint's fulfillment is detected by a pattern that identifies a valid target token following an activation within a given time window. Dynamic correlation constraints (e.g., matching a response parameter to an activation via cross-event data fields) are verified inline by mapping relational attributes across discrete event schemas. When a pattern is successfully matched or its deadline expires, \textbf{L2} emits a derived lifecycle milestone event, such as:
\begin{itemize}
  \item \texttt{ConstraintFulfillment}: \(\langle caseId,\, constraintId,\, activationId,\, ts \rangle\).
  \item \texttt{ConstraintViolation}: \(\langle caseId,\, constraintId,\, activationId,\, ts,\, reason \rangle\).
\end{itemize}
This layer remains entirely stateless regarding process state; it does not track overall case history but rather announces critical transactional events that define individual constraint statuses.

\paragraph{Process Status Layer (L3)}
The \emph{Process Status Layer (L3)} builds on the lifecycle milestones emitted by the pattern-matching layer and provides a stateful view of execution for each unique process case. Its primary role is twofold: (i) to manage and aggregate the status of all individual constraints into an overall case-level picture, and (ii) to trigger process-level actions whenever significant state changes occur. This layer consists of stateful listeners that subscribe to the fulfillment and violation event streams from \textbf{L2}. For example, a constraint instance begins in an \textsf{Idle} state. Upon receiving an \texttt{ACTIVATION} token forwarded from \textbf{L1}, the tracking routine marks the constraint as \textsf{Activated}. When it subsequently receives a structural fulfillment token from \textbf{L2}, it transitions the state to \textsf{Fulfilled}. 

Beyond tracking, this evaluation routine interprets lifecycle milestones to maintain active process state vectors. These state transitions dynamically compute allowed subsequent operations (\textbf{Task Enablement}) and trace-wide validity (\textbf{Instance Finishability}), rendering the current permissible actions transparently visible to actors or external applications. When an internal metric boundary intersects a task execution condition, a separate enforcement unit tracks sustained state periods using an operational execution construct modeled as a 6-tuple $\mathcal{R}=(\Sigma,\pi,\delta,\alpha,\rho,\gamma)$, detailed in \cref{tab:enforcement_tuple}. 

Finally, \textbf{L3} is responsible for initiating automated or corrective actions; when a state transition to \textsf{Fulfilled} or \textsf{Violated} occurs, listeners can directly trigger external services, notify target systems, or launch compensating activities, effectively closing the semantic loop between continuous physical events and business execution.

\begin{table}[t]
  \centering
  \begin{talltblr}[
      caption = {Formal Components of the Enforcement Service Tuple},
      label   = {tab:enforcement_tuple},
    ]{
      width   = \linewidth,
      colspec = {X[l,-1]X[l,-1]X[l,-1]},
      row{1}  = {font=\bfseries},
      hline{1,Z} = {1pt,solid},
      hline{2}   = {.5pt,solid},
      stretch=-1
    }
    \textbf{Element} & \textbf{Domain} & \textbf{Operational Definition} \\
    $\Sigma$ & Stream Identifier & The target high-frequency signal stream source emitted by an external sensor or physical system (e.g., \textit{temperature}). \\
    $\pi$    & Predicate Logic   & The conditional expression evaluated over payload metrics (e.g., $\mathit{value} > 80.0$). \\
    $\delta$ & Temporal Window   & The minimum continuous duration threshold required to validate a breach. \\
    $\alpha$ & Action Event      & The discrete synthetic event injected into the stream upon violation to trigger machine actuation or alert human operators. \\
    $\rho$   & Response Payload  & The key-value payload attached to the injected action event, allowing parameterized actuation (e.g., setting a target temperature or routing an alert to a specific operator). \\
    $\gamma$ & Reset Predicate   & The condition (e.g., $\neg\pi$ sustained, or a separate recovery threshold) under which an active breach is cleared, determining whether $\alpha$ fires once per breach episode or repeatedly while $\pi$ holds. \\
  \end{talltblr}
\end{table}

By establishing these systematic mappings, the execution pipeline shifts from an ad hoc tracking routine into a deterministic compilation framework. The operationalization of these abstract components within individual runtime layers is explained below for two exemplary templates:

\textbf{Hybrid Response Constraint.}
The \textsc{Hybrid Response} template requires that whenever an activation condition \(\Theta_1\) holds, a response condition \(\Theta_2\) must occur within a bounded interval, where \(I_a\) and \(I_r\) denote the activation and response windows. Its mapping proceeds as follows. At \textbf{L1}, two queries detect \(\Theta_1\) and \(\Theta_2\), emitting \texttt{ACTIVATION} and \texttt{TARGET} tokens. At \textbf{L2}, a pattern checks whether a \texttt{TARGET} follows an \texttt{ACTIVATION} within \(I_r\), producing a fulfillment event; a complementary pattern with a timer emits a violation token when no such match occurs. At \textbf{L3}, a listener updates the constraint state: it marks an \texttt{ACTIVATION} as \textsf{Activated} and transitions the instance to \textsf{Fulfilled} or \textsf{Violated} upon receiving the corresponding L2 event.

\textbf{Hybrid NotExistence Constraint.}
The \textsc{Hybrid NotExistence} template demands that a condition \(\Theta\) never occur within a time interval \(I\). Its mapping is simpler. At \textbf{L1}, a query detects whether \(\Theta\) ever becomes true and emits an event if so. At \textbf{L2}, this is converted into a constraint violation event. At \textbf{L3}, the listener sets the constraint state to \textsf{Violated}. If no such event is received before the end of \(I\), the constraint is considered fulfilled, and the listener updates the structural execution state accordingly.

\subsection{Conceptual Hybrid Constraint Mapping to CEP Logic}
To provide a generalized, implementation-independent mapping from formal hybrid declarative specifications to executable data streams, the compilation pipeline is formalized as a taxonomy matrix. \Cref{tab:cel_stl_epl_mapping} demonstrates how traditional discrete Declare templates are lifted into continuous-time STL semantics \cite{corea2025hybrid}, abstracted via intermediate Denotational Complex Event Logic (CEL) \cite{Grez2019_Formal} expressions, and operationalized into concrete Esper EPL \cite{Langhi2025} implementation syntax. This formalization ensures that declarative patterns can be systematically evaluated in real time regardless of whether their internal predicates monitor symbolic events or continuous sensor thresholds. The groups in \cref{tab:cel_stl_epl_mapping} represent the fundamental constraint families; stricter variants (e.g., alternate, chain), negated forms, and weaker forms (e.g., responded existence) are translated structurally equivalently with minor query adjustments.

Furthermore, moving beyond simple temporal control flows requires decomposing multi-perspective (MP) declarative process models into their constitutional building blocks: \textit{Activation}, \textit{Target}, \textit{Correlation}, and \textit{Temporal conditions} \cite{burattin2016conformance}. \Cref{tab:mp_declare_component_decomposition} specifies this decomposition by demonstrating how first-order data-driven predicates and cross-event data relationships ($\varphi_c(x,y)$) are managed across the runtime pipeline. 

  \begin{sidewaystable*}[p]
    \centering
    \caption{Taxonomy Matrix: Mapping Hybrid Process Specifications (STL) to Executable Rules (Esper EPL) via Denotational Complex Event Logic (CEL) Semantics}
    \label{tab:cel_stl_epl_mapping}
    \tiny
    \begin{tblr}{
        width = \linewidth,
        colspec = {
          X[l,-1,font=\bfseries]X[l,-1]X[l,2]X[l,2]X[l,2,font=\ttfamily]
        },
        row{1} = {font=\bfseries},
        hline{1,Z} = {1pt,solid},
        hline{2}   = {.5pt,solid},
        stretch=-1
      }

      Template Group & Traditional Declare LTL & Hybrid STL Semantics \cite{maler2004monitoring} & Complex Event Logic (CEL) \cite{Grez2019_Formal} Semantics & Concrete Esper EPL \cite{Langhi2025} Implementation \\

      Existence &
      $F a$ \newline\newline \textbf{Target:} $a$ &
      $F_I(\Theta)$ \newline\newline Predicate $\Theta$ must be satisfied at some point within interval $I$. &
      $(\Theta \text{ AS } x) \text{ FILTER } (x.\mathit{ts} \in I)$ \newline\newline Denotes a stream match instance where predicate $\Theta$ falls inside temporal bounds $I$. &
      INSERT INTO constraintStatus \newline
      SELECT id, '1' as name, 'TARGET' as type, timestamp \newline
      FROM GenericEvent($\Theta$) \newline
      WHERE timer:within(I sec); \\

      NotExistence &
      $G(\neg a)$ \newline\newline \textbf{Target:} $a$  &
      $\neg F_I(\Theta)$ \newline\newline Evaluates an invariant safety rule forbidding condition $\Theta$ across window $I$. &
      $\neg (\Theta \text{ AS } x) \text{ FILTER } (x.\mathit{ts} \in I)$ \newline\newline Enforces an empty set partition for condition matches over the evaluation window. &
      INSERT INTO constraintStatus \newline
      SELECT id, '1' as name, 'PERMANENT\_VIOLATION' as type, timestamp \newline
      FROM GenericEvent \newline
      WHERE $\Theta$ AND timer:within(I sec); \\

      Response &
      $G(a \rightarrow F b)$ \newline\newline \textbf{Activation:} $a$ \newline \textbf{Target:} $b$  &
      $G_{I_a}(\Theta_1 \rightarrow F_{I_r}(\Theta_2))$ \newline\newline Bounded forward temporal causation linking activation to target window $I_r$. &
      $((\Theta_1 \text{ AS } x) \cdot \text{True}^* \cdot (\Theta_2 \text{ AS } y)) \text{ FILTER } (x.\mathit{ts} \in I_a \wedge y.\mathit{ts} - x.\mathit{ts} \in I_r)$ \newline\newline Sequence concatenation where an activation matches a subsequent target within $I_r$. &
      INSERT INTO constraintStatus \newline
      SELECT b.id, '2' as name, 'FULFILLMENT' as type, b.timestamp \newline
      FROM pattern [every a=GenericEvent($\Theta_1$) \newline
      -> b=GenericEvent($\Theta_2$)] \newline
      WHERE b.timestamp - a.timestamp <= $I_r$; \\

      Chain \newline Response &
      $G(a \rightarrow X b)$ \newline\newline \textbf{Activation:} $a$ \newline \textbf{Target:} $b$  &
      $G_{I_a}(\Theta_1 \rightarrow F_{[0,\epsilon]}\Theta_2)$ \newline\newline Discretizes symbolic succession via a microscopic continuous temporal offset $\epsilon$. &
      $((\Theta_1 \text{ AS } x) \cdot (\Theta_2 \text{ AS } y)) \text{ FILTER } (x.\mathit{ts} \in I_a \wedge y.\mathit{ts} - x.\mathit{ts} \le \epsilon)$ \newline\newline Clean immediate serialization operator ($\cdot$) bounded by micro-interval $\epsilon$. &
      INSERT INTO constraintStatus \newline
      SELECT b.id, '2' as name, 'FULFILLMENT' as type, b.timestamp \newline
      FROM pattern [every a=GenericEvent($\Theta_1$) \newline
      -> b=GenericEvent($\Theta_2$)] \newline
      WHERE b.timestamp - a.timestamp <= $\epsilon$; \\

      Precedence &
      $(\neg b \,U\, a) \vee G(\neg b)$ \newline\newline \textbf{Activation:} $b$ \newline \textbf{Target:} $a$  &
      $G_{I_r}(\Theta_2 \rightarrow O_{I_a}\Theta_1)$ \newline\newline Restricts execution of target $\Theta_2$ contingent on a verified historical activation within $I_a$. &
      $((\Theta_1 \text{ AS } x) \cdot \text{True}^* \cdot (\Theta_2 \text{ AS } y)) \text{ FILTER } (y.\mathit{ts} \in I_r \wedge y.\mathit{ts} - x.\mathit{ts} \in I_a)$ \newline\newline Historical sequence calculation verifying target timestamp succeeds a valid activation. &
      INSERT INTO constraintStatus \newline
      SELECT b.id, '2' as name, 'FULFILLMENT' as type, b.timestamp \newline
      FROM pattern [every a=GenericEvent($\Theta_1$) \newline
      -> b=GenericEvent($\Theta_2$)] \newline
      WHERE b.timestamp >= a.timestamp AND (b.timestamp - a.timestamp) $\in I_a$; \\

      Not \newline Response &
      $G(a \rightarrow \neg F b)$ \newline\newline \textbf{Activation:} $a$ \newline \textbf{Target:} $b$ &
      $G_{I_a}(\Theta_1 \rightarrow \neg F_{I_r}(\Theta_2))$ \newline\newline Prohibits conflicting target events during reaction window $I_r$ following activation. &
      $\neg ((\Theta_1 \text{ AS } x) \cdot \text{True}^* \cdot (\Theta_2 \text{ AS } y)) \text{ FILTER } (x.\mathit{ts} \in I_a \wedge y.\mathit{ts} - x.\mathit{ts} \in I_r)$ \newline\newline Enforces a strict empty match partition for forbidden sequence tracking blocks. &
      INSERT INTO constraintStatus \newline
      SELECT a.id, '2' as name, 'PERMANENT\_VIOLATION' as type, a.timestamp \newline
      FROM pattern [every a=GenericEvent($\Theta_1$) \newline
      -> b=GenericEvent($\Theta_2$) and timer:interval($I_r$ sec)]; \\

    \end{tblr}
  \end{sidewaystable*}

  \begin{sidewaystable*}[p]
    \centering
    \caption{Architectural Component Decomposition: Operationalizing MP-Declare Building Blocks (Activation, Target, Correlation) through Hybrid STL, Denotational CEL Semantics, and Esper EPL Queries}
    \label{tab:mp_declare_component_decomposition}
    \tiny
    \begin{tblr}{
        width = \linewidth,
        colspec = {
          X[l,-1, font=\bfseries]X[l,-1]X[l,2]X[l,2]X[l,1, font=\ttfamily]
        },
        row{1} = {font=\bfseries},
        hline{1,Z} = {1pt,solid},
        hline{2}   = {.5pt,solid},
        stretch=-1
      }

      MP Condition Type & MP-Declare / MFOTL Semantics \cite{burattin2016conformance} & Complex Event Logic (CEL) \cite{Grez2019_Formal} Semantics & CEP Execution Role (Runtime) & Concrete Esper EPL \cite{Langhi2025} Implementation \\

      Activation \newline Condition ($\varphi_a(x)$) &
      $p_A(x) \wedge r_a(x)$ \newline\newline Unary predicate evaluating global attributes of activation event $A$ at payload position $x$. &
      $(\Theta_A \text{ AS } x) \text{ FILTER } r_a(x)$ \newline\newline Binds variable alias $x$ directly to primary event definition, filtering traces instantly. &
      \textbf{Atomic Event Layer (L1):} \newline Continuous stream queries evaluate incoming scalar thresholds in real time, forwarding an \texttt{ACTIVATION} token to \textbf{L2}. &
      INSERT INTO constraintStatus \newline
      SELECT t1.id AS id, '1' AS name, \newline
      'ACTIVATION' AS type, t1.timestamp \newline
      FROM pattern [every t1=GenericEvent( \newline
          eventType='Pay', \newline
      cast(payload('type'), String)='gold')]; \\

      Target \newline Condition ($\varphi_t(y)$) &
      $p_B(y) \wedge r_t(y)$ \newline\newline Unary predicate evaluating independent parameters restricted strictly to target event $B$. &
      $(\Theta_B \text{ AS } y) \text{ FILTER } r_t(y)$ \newline\newline Unary target filtering via an independent variable expression inside a CEL statement. &
      \textbf{Atomic Event Layer (L1):} \newline Identifies target actions and evaluates properties, forwarding a tagged \texttt{TARGET} marker to \textbf{L2}. &
      INSERT INTO constraintStatus \newline
      SELECT id, '2' as name, \newline
      'TARGET' as type, timestamp \newline
      FROM GenericEvent \newline
      WHERE eventType = 'GetDiscount' \newline
      AND cast(payload('status'), String)='valid'; \\

      Correlation \newline Condition ($\varphi_c(x,y)$) &
      $r_c(x,y)$ \newline\newline Dyadic data predicate evaluated across activation instance $x$ and target instance $y$. &
      $\dots \text{ FILTER } r_c(x,y)$ \newline\newline Evaluates cross-variable relational predicates across aliases $x$ and $y$. &
      \textbf{Constraint Status Layer (L2):} \newline Evaluates multi-event schemas. It matches activation variables against incoming target records to verify dynamic payload constraints. &
      SELECT b.id, b.name, b.timestamp \newline
      FROM pattern [every \newline
        a=constraintStatus(type='ACTIVATION') \newline
      -> b=constraintStatus(type='TARGET')] \newline
      WHERE cast(b.payload('id'), String) = \newline
      cast(a.payload('id'), String); \\

      Time \newline Condition ($\varphi_\tau$) &
      $y.\mathit{ts} - x.\mathit{ts} \in I$ \newline\newline Metric interval restriction bounding chronological distance between events. &
      $\dots \text{ FILTER } (y.\mathit{ts} - x.\mathit{ts} \in I)$ \newline\newline Expresses explicit continuous-time temporal limits across stream-derived timestamp fields. &
      \textbf{Constraint Status Layer (L2):} \newline Instantiates continuous clock-driven execution windows via explicit sliding constraints to actively emit success/timeout events. &
      SELECT b.id FROM pattern [ \newline
        every a=constraintStatus(type='ACTIVATION') \newline
        -> b=constraintStatus(type='TARGET') \newline
      where timer:within(12 hour)]; \\
    \end{tblr}
  \end{sidewaystable*}

\subsection{Example Process: Hybrid Execution in a Cooling Process}
Consider a machine that performs a chemical reaction and must remain within safe temperature limits. The process involves both continuous temperature measurements and discrete activities such as \texttt{StartCooling} and \texttt{Restart}. At runtime, the execution engine continuously processes incoming events and sensor streams.

The first constraint, \(\Phi_1 = \textsc{NotExistence}_{[0,10]}(\textit{temp}(t) > 90)\), is a \textsc{NotExistence} rule acting as a critical safety boundary. It specifies that the temperature must never remain above 90\,°C for more than 10 seconds. A violation indicates that the process has entered a dangerous state beyond its normal reactive capabilities. In this case, the engine must take immediate action, such as stopping the process and raising an alert to prevent damage (see \cref{fig:NotExistence}).
\textbf{L1} detects that the signal \texttt{temp > 90°C} remains true over the full 10-second window and emits a target event;
\textbf{L2} evaluates this and emits a \texttt{constraintStatus} event indicating a permanent violation;
\textbf{L3} consumes the event, marks the constraint instance as \textsf{Violated}, and triggers a system halt and alert.

The second constraint is a \textsc{Response},
$\Phi_2 = \textsc{Response}_{[0,10],[0,5]}(\textit{temp}(t) > 80,\, \mathsf{dis}(\textit{StartCooling}[t]))$, and
defines a reactive mechanism for prolonged high temperature. If the temperature stays above 80\,°C for a continuous 10-second interval, an obligation is triggered: a \texttt{StartCooling} activity must occur within a strict 5-s response window (see \cref{fig:response}).
\textbf{L1} detects the sustained \texttt{temp > 80°C} signal and emits an \texttt{ACTIVATION} event; it also detects the discrete \texttt{StartCooling} event and emits a \texttt{TARGET} event.
\textbf{L2} evaluates whether a \texttt{TARGET} follows the \texttt{ACTIVATION} within 5-s and issues a \texttt{ConstraintFulfillment} or \texttt{Constraint Violation} accordingly.
\textbf{L3} interprets this result, updating the constraint’s state and either allowing execution to proceed or triggering the violation response.

\begin{figure}[t]
  \centering
  \resizebox{\linewidth}{!}{
\begin{tikzpicture}[transform shape]
  \def\LOneY{3.0}
  \def\LTwoY{1.5}
  \def\LThreeY{0.0}
  \def\TimelineY{-1.0}

  \draw[timeline] (0,\TimelineY) -- (14,\TimelineY) node[right] {$t$};
  \foreach \x/\label in {0/0s, 2.5/5s, 5/10s, 7.5/15s, 10/20s, 12.5/25s} {
    \draw (\x,\TimelineY+0.1) -- (\x,\TimelineY-0.1) node[below] {\label};
  }
  \node[layer_label] at (-1, \LOneY) {L1};
  \node[layer_label] at (-1, \LTwoY) {L2};
  \node[layer_label] at (-1, \LThreeY) {L3};

  \draw[dashed, black!30, thin] (0, \LOneY+0.5) -- (13.5, \LOneY+0.5) node[right, black!60] {\scriptsize $90^\circ\text{C}$};

  \draw[thick, black] (0, \LOneY-0.3) -- (7.5, \LOneY-0.3) 
    -- (7.5, \LOneY+0.65) -- (12.5, \LOneY+0.65) 
    -- (12.5, \LOneY-0.5) -- (13.5, \LOneY-0.5);
  
  \node[left, font=\small] at (0, \LOneY-0.3) {$\text{temp}(t)$ };

  \draw[failurered, line width=4pt] (7.5, \LOneY+0.65) -- (12.5, \LOneY+0.65) node[midway, above=1mm] {$\text{temp} > 90^\circ\text{C}$};

  \draw[-Stealth, loosely dotted, thick, infogray] (12.5, \LOneY-0.5) -- (12.5, \LTwoY+0.3);
  \node[icon_failure] at (12.5, \LTwoY) (violation) {\faCircleXmark};

  \node[error_box, text width=3cm] at (12.5, \LThreeY) (case_violated) {Case: Violated \\ \small\textit{Halt Process \& Raise Alert}};
  \draw[-latex, failurered] (violation) -- (case_violated);
  \node[anno, failurered] at (10, \LOneY+0.1) {\scriptsize 10s};
\end{tikzpicture}
  }
  \caption{Example \textsc{NotExistence} constraint execution via CEP streams.}
  \label{fig:NotExistence}
\end{figure}

\begin{figure}[t]
  \centering
  \resizebox{\linewidth}{!}{
\begin{tikzpicture}[transform shape]
  \def\LOneY{3.0}
  \def\LTwoY{1.5}
  \def\LThreeY{0.0}
  \def\TimelineY{-1.0}

  \draw[timeline] (0,\TimelineY) -- (14,\TimelineY) node[right] {$t$};
  \foreach \x/\label in {0/0s, 2.5/5s, 5/10s, 7.5/15s, 10/20s, 12.5/25s} {
    \draw (\x,\TimelineY+0.1) -- (\x,\TimelineY-0.1) node[below] {\label};
  }
  \node[layer_label] at (-1, \LOneY) {L1};
  \node[layer_label] at (-1, \LTwoY) {L2};
  \node[layer_label] at (-1, \LThreeY) {L3};

  \draw[dashed, black!30, thin] (0, \LOneY+0.4) -- (13.5, \LOneY+0.4) node[right, black!60] {\scriptsize $80^\circ\text{C}$};

  \draw[thick, black] (-.2, \LOneY+.6) -- (5.625, \LOneY+0.6) 
    -- (5.625, \LOneY-0.4) -- (13.5, \LOneY-0.4);
  
  \node[left, font=\small] at (-.2, \LOneY+0.6) {$\text{temp}(t)$ };

  \draw[riskycondition, line width=4pt] (0, \LOneY+0.6) -- (5.0, \LOneY+0.6) node[midway, above=1mm] {$\text{temp} > 80^\circ\text{C}$};
  
  \node[event] at (5.625, \LOneY-0.4) (cooling_event) {};
  \node[above right=1mm of cooling_event, font=\small, color=signalblue] {StartCooling};

  \draw[-latex, dashed, infogray, thick] (5.0, \LOneY+0.45) -- (5.0, \LTwoY+0.3);
  \node[riskycondition, font=\bfseries] at (5.0, \LTwoY-.35) {\faCircleExclamation};
  \node[anno, riskycondition, font=\bfseries] at (5.0, \LTwoY-0.75) {ACTIVATION};

  \node[zone, minimum width=2.75cm, anchor=west, minimum height=0.7cm] at (5.0, \LTwoY) {};
  \node[anno] at (6.4, \LTwoY) {Response Window \\ \scriptsize 5s};

  \node[icon_success] at (5.625, \LTwoY-.35) (fulfillment) {\faCircleCheck};
  \draw[-latex, dashed, signalblue, thick] (cooling_event) -- (fulfillment);
  
  \node[icon_failure] at (7.5, \LTwoY-.35) (violation) {\faCircleXmark};

  \node[ok_box] at (5.625, \LThreeY) (case_ok) {Case: OK \\ \small\textit{Continue Process}};
  \draw[-latex, successgreen] (fulfillment) -- (case_ok);

  \node[error_box, text width=3.5cm] at (9.5, \LThreeY) (case_violated) {Case: Violated \\ \small\textit{Auto-trigger cooling}};
  \draw[-latex, failurered] (violation) -- (case_violated);

  \node[anno, riskycondition] at (2.5, \LOneY+0.1) {\scriptsize 10s};
\end{tikzpicture}
  }
  \caption{Example \textsc{Response} constraint execution via CEP streams.}
  \label{fig:response}
\end{figure}

\begin{figure}[h]
  \centering
  \resizebox{\linewidth}{!}{
\begin{tikzpicture}[transform shape]
  \def\LOneY{3.0}
  \def\LTwoY{1.5}
  \def\LThreeY{0.0}
  \def\TimelineY{-1.0}
  \draw[timeline] (0,\TimelineY) -- (14,\TimelineY) node[right] {$t$};
  \foreach \x/\label in {0/0s, 2.5/10s, 5/20s, 7.5/30s, 10/40s, 12.5/50s} {
    \draw (\x,\TimelineY+0.1) -- (\x,\TimelineY-0.1) node[below] {\label};
  }
  \node[layer_label] at (-1, \LOneY) {L1};
  \node[layer_label] at (-1, \LTwoY) {L2};
  \node[layer_label] at (-1, \LThreeY) {L3};
  \draw[dashed, black!30, thin] (0, \LOneY-0.4) -- (13.5, \LOneY-0.4) node[right, black!60] {\scriptsize $50^\circ\text{C}$};

  \draw[thick, black] (0, \LOneY+0.5) -- (2.5, \LOneY+0.5) 
    -- (2.5, \LOneY-0.7) -- (13.5, \LOneY-0.7);
  
  \node[left, font=\small] at (0, \LOneY+0.5) {$\text{temp}(t)$ };

  \draw[safecondition, line width=4pt] (2.5, \LOneY-0.7) -- (7.5, \LOneY-0.7) node[pos=.3, above=3mm] {$\text{temp} < 50^\circ\text{C}$};
  
  \node[event] at (6.0, \LOneY-0.7) (attempt1) {};
  \node[above=2mm of attempt1, text width=2cm, align=center, font=\small, color=signalblue] {Restart \\ \scriptsize(Attempted)};
  
  \node[event] at (10.0, \LOneY-0.7) (attempt2) {};
  \node[above=2mm of attempt2, font=\small, color=signalblue] {Restart \scriptsize(Successful)};

  \node[zone, minimum width=5cm, anchor=west, minimum height=0.7cm, dashed] at (2.5, \LTwoY) {};
  \node[anno] at (5.0, \LTwoY) {Enabling Countdown};
  
  \draw[-latex, dashed, failurered] (6.0, \LOneY-0.9) -- (6.0, \LTwoY+0.3);
  \node[icon_failure] at (6.0, \LTwoY-.35) (blocked) {\faCircleXmark};
  \node[anno_failure, below left=-1.5mm and 0mm of blocked] {BLOCKED};
  
  \draw[-latex, dashed, infogray] (7.5, \LOneY-0.5) -- (7.5, \LTwoY+0.3);
  \node[icon_success] at (7.5, \LTwoY-.35) (enabled) {\faCircleCheck};
  \node[anno_success, below=-1.5mm of enabled] {ENABLED};

  \node[error_box, text width=3cm] at (6.0, \LThreeY-0.2) (action_blocked) {Case: Blocked \\ \small\textit{Action 'Restart' not allowed}};
  \draw[-latex, failurered] (blocked) -- (action_blocked);
  
  \node[ok_box] at (10.0, \LThreeY-0.2) (case_ok) {Case: OK \\\small\textit{Process Resumed}};
  \draw[-latex, successgreen] (attempt2) -- (case_ok.north);

  \node[anno, successgreen] at (3.75, \LOneY-.95) {\scriptsize 20s};
\end{tikzpicture}
  }
  \caption{Example \textsc{Precedence} constraint execution via CEP streams.}
  \label{fig:precedence}
\end{figure}

The third constraint is a \textsc{Precedence}, $\Phi_3 = \textsc{Precedence}_{[0,20],[0,20]}(\textit{temp}(t) < 50,\, \mathsf{dis}(\textit{Restart}[t]))$, which enforces a precondition for the sensitive \texttt{Restart} activity (see \cref{fig:precedence}). The activity \texttt{Restart} is permitted only if the system has been in a cool, stable state (temperature below 50\,°C) for a continuous window of at least 20 seconds. Here, \textbf{L1} generates a sustained state activation token once the 20-second cooling condition is satisfied, allowing \textbf{L2} to validate any subsequent \texttt{Restart} attempts against this active window. Until this duration condition is met, the engine indicates that \texttt{Restart} cannot yet occur, preventing unsafe execution.
\textbf{L1} receives the \texttt{temp < 50°C} stream and two \texttt{Restart} attempts;
\textbf{L2} determines that the first attempt violates the required 20-second stability window and emits a \texttt{ConstraintViolation}; once the 20 seconds of stability have elapsed, the condition becomes enabled;
\textbf{L3} consumes the violation event, triggers compensatory action, and finally accepts the second valid \texttt{Restart} attempt, allowing the process to resume.

This example illustrates how declarative templates evolve from descriptive constraints into prescriptive runtime logic. By continuously interpreting hybrid specifications and executing associated actions, the system bridges the gap between modeling and operational control in cyber-physical process environments.

\section{Implementation}\label{sec:implementation}
We developed a prototype that enforces hybrid declarative constraints via CEP rules.\footnote{See \url{https://github.com/LeoPoss/quarkusCEP/tree/STL_clean} for the code repository and a short guided introduction at \url{https://youtu.be/7FwQxgyozcM}.} In contrast to existing approaches, where constraint conditions are purely discrete (e.g., the occurrence of an activity), hybrid constraints may involve continuous predicates evaluated over time windows. Our architecture is implemented as a layered, event-driven runtime built on top of the Esper CEP engine. Esper provides a high-performance environment for evaluating continuous queries over streaming data using its Event Processing Language (EPL), which is well-suited for operationalizing declarative process semantics.

\textbf{L1: Atomic Event Detection.}
This layer ingests and normalizes runtime data into typed event streams. Both discrete atomic events and continuous signals are represented using the \texttt{GenericEvent} class, which contains an \texttt{eventType} and a payload map specifying the type (Signal/Task) and either the measured value (e.g., \texttt{temp=100}) or task-related metadata (e.g., \texttt{userId=3}).
The final step in \textbf{L1} transforms these raw events into semantic atomic events in a unified \texttt{constraintStatus} stream, which serves as the input for \textbf{L2} and contains all activations and targets relevant for constraint monitoring, as shown in \cref{lst:l1}.
The \texttt{constraintStatus} stream is instantiated upon the first insertion and stores the event ID, the constraint name, the event type (\texttt{ACTIVATION} or \texttt{TARGET}), a timestamp, and the payload used for correlation at \textbf{L2}.

\begin{lstlisting}[language=EPL,style=epl,caption={Detection of \texttt{ACTIVATION} an \texttt{TARGET} events of constraints (\textbf{L1}).},label={lst:l1}]
--- NOT EXISTENCE TARGET for temperature > 90 without a parameter timer
INSERT INTO constraintStatus
SELECT id, '1' as name, 'TARGET' as type, timestamp as timestamp, cast(payload('temp'), double) as temp
FROM GenericEvent
WHERE eventType = 'Temp' AND cast(payload('temp'), double) > 90

--- RESPONSE ACTIVATION where temp > 80 for 10 seconds
INSERT INTO constraintStatus
SELECT t1.id AS id, '2' AS name, 'ACTIVATION' AS type, t1.timestamp AS timestamp, cast(t1.payload('temp'), double) AS temp
FROM pattern [every t1 = GenericEvent(eventType = 'Temp', cast(payload('temp'), double) > 80) -> (timer:interval(10 sec) and not GenericEvent(eventType = 'Temp', cast(payload('temp'), double) <= 80))]
    \end{lstlisting}
    
\begin{lstlisting}[language=EPL,style=epl,caption={Query for detecting fulfillment patterns (\textbf{L2}).},label={lst:L2ful}]
SELECT b.id, b.name, b.type, b.timestamp as timestamp
FROM PATTERN [every a=constraintStatus(type='ACTIVATION', name='2') -> b=constraintStatus(type='TARGET', name='2')]
    \end{lstlisting}

\begin{lstlisting}[language=EPL,style=epl,caption={Query for detecting violation patterns (\textbf{L2}).},label={lst:L2vio}]
INSERT INTO constraintStatus
SELECT '2' as name, 'PERMANENT_VIOLATION' as type, a.timestamp as timestamp
FROM PATTERN [
    a=constraintStatus(type='ACTIVATION', name='2')
    -> (timer:interval(5 sec) and not b=constraintStatus(type='TARGET', name='2'))
]
    \end{lstlisting}

\textbf{L2: Constraint-Level Pattern Matching.}
The second layer consumes the \texttt{constraintStatus} stream and uses EPL pattern statements to detect fulfillments and violations, emitting standardized lifecycle events.
A fulfillment is implemented as a pattern query over the \texttt{constraintStatus} stream (see \cref{lst:L2ful}), while violations are detected using negative patterns with timeouts (see \cref{lst:L2vio}, where a \textsc{Response} must occur within 5 seconds of its activation).

\textbf{L3: Process-Level State Management.}
The output of \textbf{L2} forms the \texttt{constraintStatus} layer at the process level, consisting of lifecycle events that represent the current state of every constraint instance, thereby defining the system’s overall state in relation to the frontend. We use Esper’s listener mechanisms to consume these events and update the internal state accordingly. Additional logic can be embedded to respond to state changes, for example, triggering a compensation service whenever any constraint becomes permanently violated. The corresponding subscriber implementation invokes an external service directly from this layer, thereby closing the loop between declarative semantics and operational execution.

\paragraph{Integration and Execution Model} Declarative execution differs fundamentally from traditional imperative process enactment (e.g., token-passing in BPMN). Imperative engines dictate the sequence of operations, advancing tokens along explicit control-flow paths and executing service tasks directly. A declarative engine does not prescribe a specific execution path. Instead, it acts as a dynamic state-space constraint filter. The primary role of our CEP-driven runtime is to evaluate ongoing event-signal configurations to compute two real-time state vectors: (i) \emph{Task Enablement}, which defines the exact subset of activities that an operator is permitted to perform, and (ii) \emph{Instance Finishability}, which determines whether the cumulative process trace satisfies all mandatory constraints, allowing the instance to terminate safely. Our framework bridges this gap through restrictive gatekeeping and active operational automation, leveraging \textbf{L3} listeners. When an STL threshold activation or a discrete task transition occurs, the process layer intercepts the lifecycle state change and translates it into an event-driven trigger. This allows the system to invoke external APIs or inject automated activity updates, effectively matching the functional execution capability of imperative service tasks while preserving the structural flexibility of a declarative model.

The three layers operate as a continuous, reactive dataflow pipeline: events are ingested and normalized in \textbf{L1}, complex patterns are detected in \textbf{L2}, and lifecycle events are interpreted to manage process state and trigger actions in \textbf{L3}.
All logic is expressed as continuous EPL queries and event-driven callbacks, enabling low-latency monitoring and control.
For actual process execution, we also track the current tasks and signals to provide users with feedback on which tasks are possible or restricted.
This is done in the spirit of \cite{Ackermann2018_Execution}, but fully separated from the CEP logic: decisions are based exclusively on the current constraints and the observed trace.
We distinguish two checks: \textit{(i)} whether the process instance can be completed, i.e., none of the constraints is temporarily or permanently violated, and \textit{(ii)} whether executing a given task would induce a violation.
This allows the system to inform users if a task is entirely forbidden or only allowed under specific payload conditions (e.g., \texttt{RestartMachine} may not be executed by \texttt{userId=3} due to the target condition of a \textsc{NotResponse} constraint).

\paragraph{Limitations and Discussion}
While the three-layer architecture handles standard multi-perspective Declare templates, translating the full spectrum of STL and $\text{LTL}_f$ into CEP syntax has three main technical limits:

\begin{enumerate}
  \item \textbf{Arbitrary Operator Nesting Constraints:} Temporal logics permit arbitrary nesting of operators (e.g., $\mathsf{F}(\mathsf{G}(\mathsf{F}_{[0,d]} \Theta))$). Esper's EPL uses a strict left-to-right sequential grammar via the linear-time developer operator (\texttt{->}). Deeply nested temporal paths cannot always be expressed as a single continuous query. Executing them requires breaking the expression into a cascade of intermediate streams, in which \textbf{L2} feeds the complex-event output of a child sub-query back as an atomic trigger token into a parent statement. This is structurally achievable but adds configuration overhead and tracking complexity.
  \item \textbf{Continuous Sampling and Signal Aliasing:} STL evaluates properties over true continuous-time real-valued functions. A discrete CEP engine processes inputs as discrete, chronologically ordered packets. If a physical signal fluctuates rapidly between two sensor ticks (e.g., a microsecond spike above a safety threshold), the engine cannot detect the transition unless the ingestion layer adds continuous interpolation or fine-grained edge sampling. Evaluation precision is therefore bounded by the sampling frequency of the physical IoT infrastructure.
  \item \textbf{State-Space Accumulation in Unbounded Windows:} Templates with long or unbounded metric intervals (e.g., tracking whether an event happens "anytime in the next three months") force the CEP engine to retain event tokens in memory for extended durations. For high-volume streams, this degrades performance and increases garbage collection. Our architecture mitigates this by focusing on metric temporal logic fragments with bounded sliding operational windows ($\Delta t$), which keeps memory profiles predictable and flat.
\end{enumerate}

\section{Evaluation}\label{sec:evaluation}

\begin{figure}[t!]
  \centering
  \includegraphics[width=\linewidth]{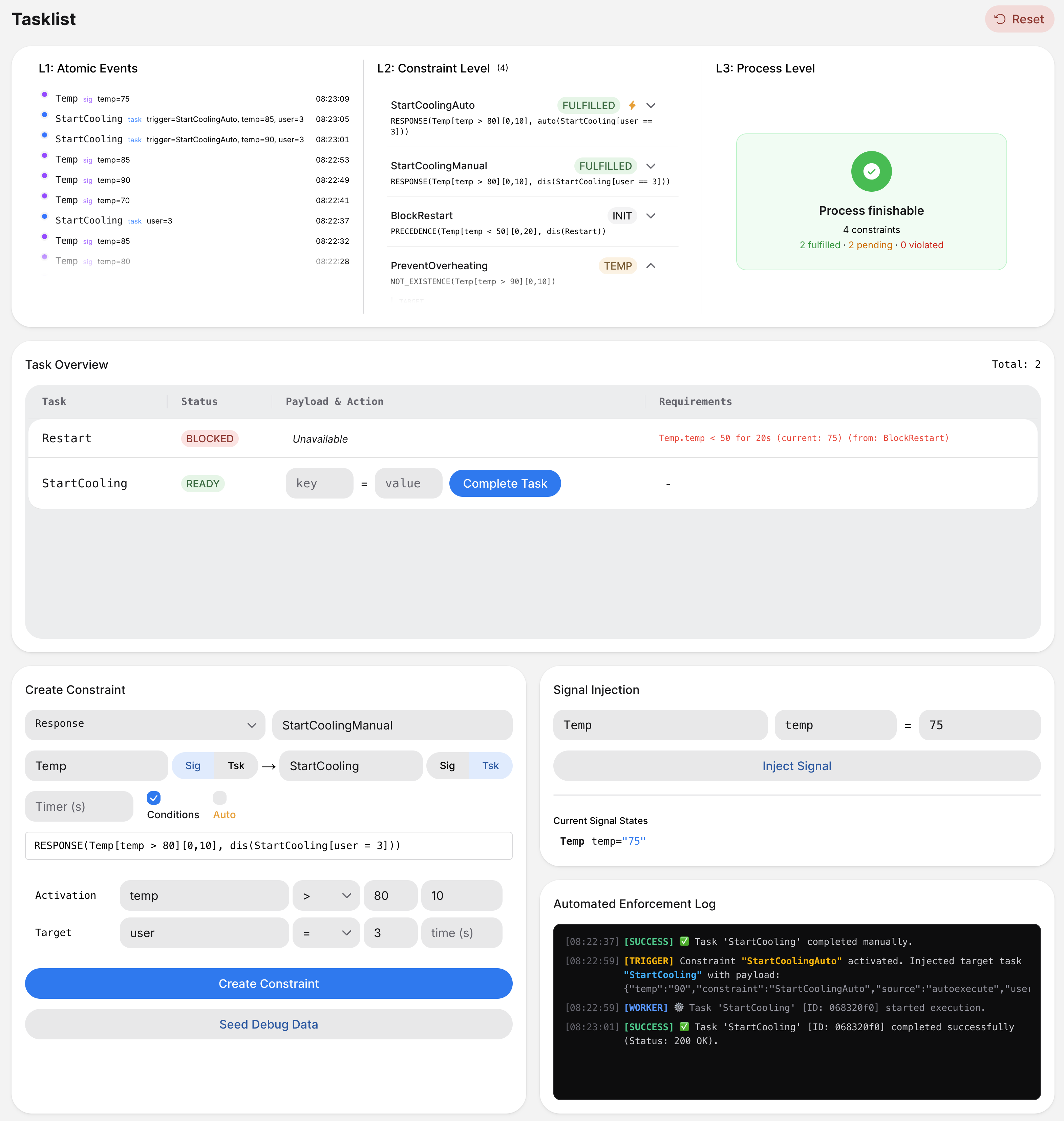}
  \caption{Overview of implemented prototype frontend}
  \label{fig:impl}
  %\vspace{-.6cm}
\end{figure}
The enforcement mechanisms of the architecture are demonstrated through an interactive process dashboard (cf.~\cref{fig:impl}). The interface comprises a signal injection terminal for sending continuous telemetry samples and a task list that polls the \texttt{AnalyzerService} to determine which activities are currently executable. Together, these components exercise the two execution modes of the runtime engine:

\begin{enumerate}
  \item The \texttt{AnalyzerService} evaluates each known task against the current constraint states and trace history. Only tasks that would not violate any active constraint are presented with an execute button; tasks blocked by unmet conditions (for instance, \texttt{Restart} while the temperature signal has not yet maintained the 20-second stability window required by $\Phi_3$) are shown without an execution option, with the specific unmet requirements (e.g., \texttt{temp < 50} for the required duration) displayed in a tooltip.\footnote{This view is similar to existing implementations of tasks lists, and could be further adjusted in future iterations.}
  \item Constraints can be configured with an auto-execute flag in the constraint creation interface. When the corresponding activation condition is detected (for example, a sensor push driving the telemetry stream above 80\,°C for 10 seconds), the engine's listener triggers the \texttt{TaskExecutorService}, which actively executes the target task and emits its completion event into the stream. The dashboard reflects this loop in real time: the auto-executed task moves from the queue to the process trace without operator intervention, confirming that the architecture drives the process lifecycle directly in response to signal changes.
\end{enumerate}

\paragraph{Expressiveness Analysis}
To evaluate the expressiveness of the proposed architecture, we examined constraints along two orthogonal dimensions: \emph{Constraint Structure} (the complexity and arity of data predicates) and \emph{Temporal Directionality} (the relative chronological ordering of event sequences). To validate that the CEP-driven engine preserves the expressiveness of the hybrid declarative paradigm, we map the cross-product of these dimensions to the execution rules in the taxonomy matrix of~\cref{tab:cel_stl_epl_mapping} and the component decomposition of~\cref{tab:mp_declare_component_decomposition}. This mapping covers all behavioral categories:

\begin{enumerate}
  \item \textbf{Atemporal Unary Structures:} Templates such as \textsc{Existence} and \textsc{NotExistence} (see~\cref{tab:cel_stl_epl_mapping}). The engine deploys an isolated, window-bounded sliding filter statement ($\Theta$) at \textbf{L1}. Execution depends on the presence or absence of an atomic token match within interval $I$, requiring no sequence-state persistence at \textbf{L2}.
  \item \textbf{Forward-Looking Binary Structures:} Standard causal templates such as \textsc{Response} and \textsc{Chain Response} (see~\cref{tab:cel_stl_epl_mapping}). \textbf{L1} captures independent activation ($\Theta_1$) and target ($\Theta_2$) components as separate stream entries. \textbf{L2} evaluates forward temporal intervals using the linear-time developer operator (\texttt{->}) with sliding time-bounds (\texttt{timer:within(I)}). If the target condition arrives within the window, the obligation is resolved; otherwise, a timeout sub-query emits a permanent violation token.
  \item \textbf{Backward-Looking Historical Structures:} Represented by the \textsc{Precedence} template. Verifying historical dependencies over real-time streams is handled by converting the forward event-driven sequence into a backward-looking filter. As the taxonomy shows, when target event $\Theta_2$ arrives, the engine compares its timestamp against cached historical records. If an activation event ($\Theta_1$) exists within the valid lookback window $I_a$, execution is approved; otherwise, it is blocked.
  \item \textbf{Multi-Perspective Data-Correlated Structures:} Detailed in~\cref{tab:mp_declare_component_decomposition}. Beyond control-flow sequencing, our engine handles data dependencies by decomposing the multi-perspective template into its components: the unary activation filter $\varphi_a(x)$, the unary target filter $\varphi_t(y)$, and the dyadic correlation constraint $\varphi_c(x,y)$. \textbf{L1} filters streams using the unary properties; \textbf{L2} uses relational join conditions to bind attributes across different event schemas (e.g., matching a \texttt{caseId}).
\end{enumerate}

Validating the runtime pipeline against this taxonomy shows that the translation from STL to EPL introduces no loss of expressiveness. Every logical connective, temporal operator, and variable data bind maps to a deterministic execution script inside the streaming engine.

\paragraph{Scalability Analysis}
To evaluate the runtime scalability and resource efficiency of the architecture under high-frequency IoT signal volumes, we subjected the prototype engine to an intensive synthetic stress test. The entire execution suite was isolated within a Docker container limited strictly to $2\text{~GB}$ of RAM and $2$ CPU cores. This resource ceiling simulates the processing constraints of industrial edge gateways, such as Raspberry Pi devices, deployed locally for shop-floor process control in Industry 4.0 environments.

The test workload used a synthetic stream. The evaluation model consisted of 10 declarative constraints deployed simultaneously: five \textsc{Response} templates and five \textsc{Precedence} templates, each with randomized activation/target conditions and varying time windows. From the Esper engine's perspective, continuous signal events and discrete task events are structurally identical \texttt{GenericEvent} objects that differ only in their \texttt{eventType}; the EPL predicate (\texttt{cast(payload(\dots), double)}) runs the same bytecode whether the payload holds a temperature reading or a task attribute. A broader evaluation of templates for discrete-event data can be found in~\cite{Poss2025_Synergistic}. We subjected this configuration to five ingestion profiles: 100, 1{,}000, 10{,}000, 50{,}000, and 100{,}000 events per second. To gather robust latency samples at each rate, the observation window is scaled inversely to the event rate: the 100~EPS profile ran for 600~s ($\approx$60k events), 1{,}000~EPS for 60~s ($\approx$60k events), and the 10{,}000--100{,}000~EPS profiles for 30~s each ($\approx$300k--3M events), with a minimum warm-up of 30~s across all configurations. Each profile was executed over $n=5$ independent runs to compute averages ($\mu$) and standard deviations ($\sigma$).

We define the evaluation metric as follows: \emph{Processing Latency} is the wall-clock time ($\Delta t$) from when a stream element enters the Atomic Layer (\textbf{L1}; measured at \texttt{System.nanoTime()} during \texttt{GenericEvent} ingestion) until the corresponding Esper rule fires at the Constraint Layer (\textbf{L2}) and updates the constraint status to \textsc{Fulfilled} or \textsc{Temporary Violation}. Latency samples triggered by timer-based \textsc{Permanent Violation} rules are excluded from the measurement because those rules fire after a configurable timeout window (\texttt{withinPeriod}) whose duration is part of the constraint semantics, not of the processing overhead. Existing hybrid declarative frameworks operate as asynchronous, offline, post hoc verification on static files; they lack an active, event-driven runtime pipeline, an ingestion mechanism, or a live execution loop. Our evaluation, therefore, focuses on demonstrating that the overhead of embedding real-time signal analysis remains within the operational bounds of edge control loops.

\begin{table}[t]
  \centering
  \begin{talltblr}[
      caption = {Performance and resource utilization metrics for scalability scenarios.},
      label = {tab:performance_metrics_scal_normal},
    ]{
      colspec = {X[l,-1]*{9}{X[r,-1]}},
      row{1-Z} = {font=\tiny},
      row{1,2} = {font=\tiny\bfseries, c, m},
      hline{1,Z} = {1pt, solid},
      hline{3} = {0.5pt, solid},
      stretch = 0,
      colsep = 3pt,
      rowsep = 2.5pt
    }
    \SetCell[r=2]{c} {Target Load}
    & \SetCell[c=2]{c} CPU (\%) & & \SetCell[c=2]{c} Memory (MB) & & \SetCell[c=3]{c} Latency (ms) & & & \SetCell[r=2]{c} {Actual Throughput\\(events/s)} \\
    \cmidrule[lr]{2-3} \cmidrule[lr]{4-5} \cmidrule[lr]{6-8}
    & $\mu$ & $\max$ & $\mu$ & $\max$ & $\mu$ & $P_{90}$ & $P_{99}$ & \\
    100 & 0.9 $\pm$ 0.1 & 19.7 $\pm$ 2.2 & 434 $\pm$ 9 & 502 $\pm$ 30 & 0.1 $\pm$ 0.0 & 0.2 $\pm$ 0.0 & 0.4 $\pm$ 0.1 & 100.0 \\
    1,000 & 6.8 $\pm$ 0.4 & 35.3 $\pm$ 2.7 & 436 $\pm$ 19 & 514 $\pm$ 12 & 0.1 $\pm$ 0.0 & 0.2 $\pm$ 0.0 & 0.5 $\pm$ 0.1 & 998.9 \\
    10,000 & 14.4 $\pm$ 0.7 & 100.9 $\pm$ 17.2 & 489 $\pm$ 5 & 518 $\pm$ 4 & 0.1 $\pm$ 0.0 & 0.2 $\pm$ 0.0 & 0.7 $\pm$ 0.0 & 9991.8 \\
    50,000 & 66.9 $\pm$ 7.7 & 203.4 $\pm$ 3.2 & 733 $\pm$ 73 & 806 $\pm$ 91 & 2.4 $\pm$ 1.2 & 1.1 $\pm$ 0.6 & 73.3 $\pm$ 38.3 & 49882.4 \\
    100,000 & 192.1 $\pm$ 0.9 & 208.1 $\pm$ 5.2 & 1586 $\pm$ 11 & 2048 $\pm$ 0 & 39.3 $\pm$ 7.6 & 94.5 $\pm$ 3.8 & 442.9 $\pm$ 97.1 & 92513.6 \\
  \end{talltblr}
\end{table}

\cref{tab:performance_metrics_scal_normal} reports resource utilization and latency measurements across the five load scenarios.
The results show that the three-layer architecture scales efficiently for realistic IoT workloads.
At event rates from 100 to 10{,}000~EPS, average CPU utilization scaled linearly from 0.9\% to 14.4\%, while memory consumption remained stable between 434 and 489~MB. Median end-to-end processing latency remained below 0.1~ms, with $P_{99}$ values under 0.5~ms, confirming sub-millisecond response times consistent with linear scaling.
At 50{,}000~EPS, CPU utilization rose to 66.9\% with peak usage of 203.4\%, indicating the onset of saturation on the 2-core configuration. Median latency remained low at 0.01~ms, though tail latency increased to $P_{99}=73.3$~ms. At the stress ceiling of 100{,}000~EPS, the system averaged 192.1\% CPU\footnote{CPU load is reported per core, so two cores equals maximum possible load of 200\%.}, memory grew to 1{,}586~MB approaching the 2~GB budget, and median latency rose to 5.9~ms with $P_{99}=442.9$~ms.\footnote{As shown in~\cite{Poss2025_Synergistic}, JVM garbage collection occurs as long as sufficient headroom exists above the Esper and Quarkus baseline consumption.}

\begin{figure}[h]
  \centering
  \includegraphics[width=\linewidth]{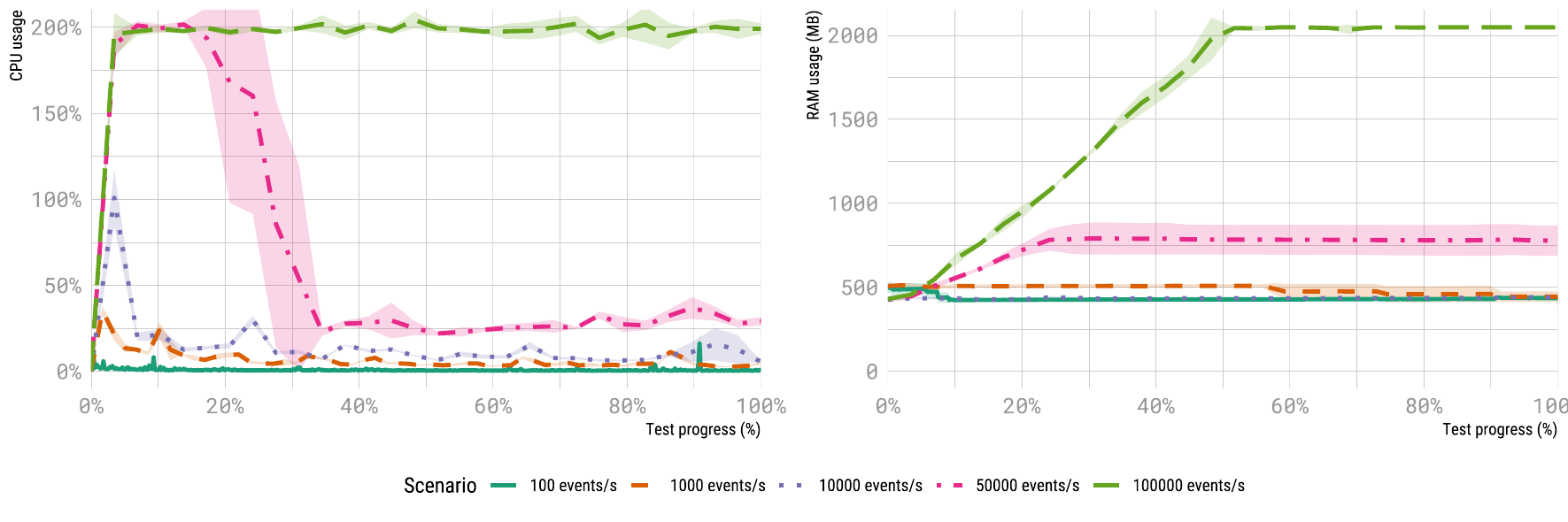}
  \caption{CPU and memory utilization across event rates. Shaded bands show 95\% confidence intervals across $n=5$ runs.}
  \label{fig:p1}
\end{figure}

\begin{figure}[h]
  \centering
  \includegraphics[width=\linewidth]{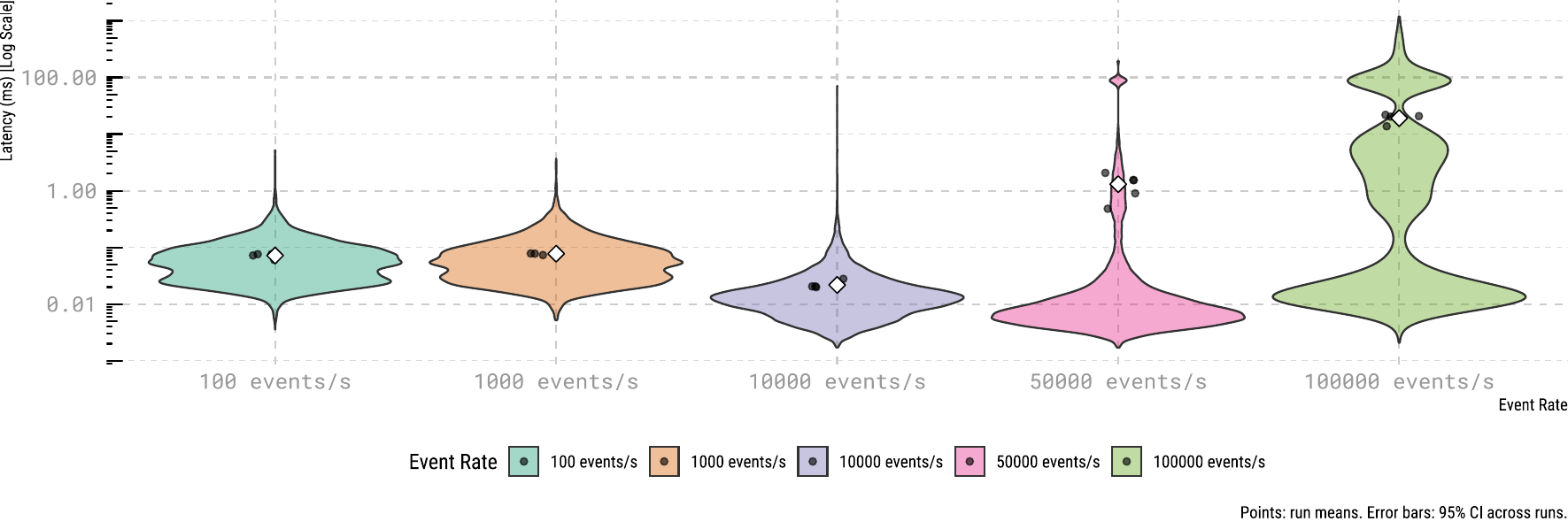}
  \caption{End-to-end processing latency by event rate (logarithmic x-axis). Violins show the full latency distribution for each rate; white diamonds mark cross-run means with 95\% confidence intervals; colored points show individual run means.}
  \label{fig:p2}
\end{figure}
\cref{fig:p1} illustrates the linear scaling behavior of resource consumption as a function of event rate, with confidence intervals derived from $n=5$ independent runs per rate. \cref{fig:p2} shows the latency distribution across event rates, with sub-millisecond response at typical IoT frequencies and graceful degradation under saturation.
These results confirm that the layered CEP architecture introduces negligible overhead while meeting a critical requirement for IoT process control: genuine real-time constraint evaluation and enforcement.
Overall, the evaluation confirms that hybrid declarative process execution is feasible on edge devices, addressing the core research gap of moving from post-hoc conformance checking to active runtime enforcement.

To validate these results on actual IoT hardware, we additionally deployed the same backend on a Raspberry Pi~5 Model~B (8~GB RAM). Using the identical constraint configuration and workload generator, we ramped the event rate from 5{,}000 to 50{,}000~EPS in 5{,}000~EPS increments (30~s per step). The Pi sustained end-to-end latencies between 0.01 and 0.67~ms across all rates. Event delivery remained above 99.7\% even at 50{,}000~EPS, indicating that the architecture had not yet reached its saturation point on the Pi~5. These results confirm that the CEP-based execution engine imposes negligible overhead on IoT-class hardware and can sustain throughput far exceeding typical industrial sensor frequencies without requiring cloud offloading or distributed infrastructure.

\section{Related Work}\label{sec:relatedwork}
BPM has focused on modeling and monitoring processes using discrete-event data, capturing sequences of activities that represent process executions \cite{dumas2018fundamentals}. Declarative approaches, such as Declare, offer a flexible alternative by specifying constraints instead of explicit control flows, which is beneficial in dynamic environments \cite{pesic2007declare,burattin2016conformance}. Classical declarative models and execution approaches \cite{Ackermann2018_Execution,KappelSSAJ19} address discrete-event traces and do not incorporate continuous sensor data, which is increasingly relevant in cyber-physical and data-rich domains \cite{bazan2022industry,janiesch2020iotbpm}. Recent work has advanced the integration of continuous data and discrete events by utilizing hybrid declarative models. Corea et al. \cite{corea2025hybrid} extended declarative process models by adopting STL, originally developed for CPS, allowing temporal properties over real-valued signals to be formally expressed and verified. This enables conformance checking of hybrid traces combining boolean discrete events and continuous predicates. This approach primarily targets offline monitoring and post-hoc conformance, rather than real-time execution of hybrid specifications.

CEP systems have demonstrated potential for runtime analysis and reaction over large-scale real-time event streams, which suits the dynamic requirements of BPM in sensor-rich environments \cite{ruhkamp2021execution,wu2006high}. State-of-the-art CEP frameworks support declarative pattern specifications and efficiently evaluate complex event patterns \cite{ziehm2024bridging}. The authors in \cite{ruhkamp2021execution,Poss2025_Synergistic} propose a CEP-based monitoring architecture that integrates multi-perspective declarative models, enabling real-time monitoring of both sensor and event data. STL has a well-established role in specifying and monitoring continuous signals within CPS \cite{maler2004monitoring,ahmad2021program}. Recent advancements extend its use to runtime enforcement, enabling proactive control based on continuous signal behavior. Ahmad and Jeannin \cite{ahmad2021program} introduced Signal Temporal Dynamic Logic (STdL), which enhances the expressiveness of reasoning about temporal properties over hybrid system executions.

While recent work has advanced hybrid specification and CEP-based monitoring, active execution of hybrid declarative models remains an open challenge. We integrate STL-inspired predicates into CEP engines to enable runtime enforcement of hybrid constraints.

\section{Conclusion}\label{sec:outlook}
This paper bridges the gap between hybrid declarative process specifications and operational control by introducing a CEP-based execution architecture for STL-enriched process models. Traditional declarative approaches have largely been limited to monitoring and post hoc conformance checking of hybrid traces, whereas our three-layer architecture enables real-time enforcement of constraints on both discrete events and continuous signals. By transforming STL-inspired predicates into CEP patterns, the system actively triggers activities and enforces process boundaries based on sensor behavior, extending declarative execution from purely discrete scenarios into IoT environments. Our evaluation demonstrates the practical feasibility of the approach on resource-constrained edge devices, achieving sub-millisecond latency at typical IoT event rates while maintaining stable resource consumption. This enables genuine real-time constraint evaluation in settings where processes must react to continuous sensor streams. The architecture successfully handles hybrid Declare templates, including unary, binary, and correlation-based constraints with both forward and backward temporal directionality.

Future work will extend this architecture toward autonomous process control by leveraging STL robustness for graded, non-binary enforcement and by integrating machine learning techniques for predictive adaptation in distributed IoT/CPS environments.

%%===========================================================================================%%
%% If you are submitting to one of the Nature Portfolio journals, using the eJP submission   %%
%% system, please include the references within the manuscript file itself. You may do this  %%
%% by copying the reference list from your .bbl file, paste it into the main manuscript .tex %%
%% file, and delete the associated \verb+\bibliography+ commands.                            %%
%%===========================================================================================%%

\bibliography{sn-bibliography}% common bib file
%% if required, the content of .bbl file can be included here once bbl is generated
%%\input sn-article.bbl

\end{document}